\documentclass{article}

\usepackage{amsmath}
\usepackage{amssymb}
\usepackage{xspace}
\usepackage{tikz}

\usepackage{floatrow}

\usepackage{enumerate}

\newtheorem{theorem}{Theorem}[section]
\newtheorem{lemma}[theorem]{Lemma}
\newtheorem{proposition}[theorem]{Proposition}

\newtheorem{conclusion}[theorem]{Conclusion}
\newtheorem{assumption}[theorem]{Assumption}
\newtheorem{proof}{Proof}

\usetikzlibrary{decorations.markings}

\newcommand{\true}{\texttt{true}}
\newcommand{\false}{\texttt{false}}

\def\endproof{\hfill$\Box$ \par}

\newcommand \dalpha {3/2}

\newcommand \ddelta {0.3}
\newcommand \tileborder {
\draw[loosely dotted] 
   (-1.5, -1.5)
-- (-1.5,  1.9)
-- ( 1.9,  1.9)
-- ( 1.9, -1.5)
-- cycle;}

\newcommand \pA {-1.5}
\newcommand \pB {-0.8}
\newcommand \pC {-0.1}
\newcommand \pD {0.2}
\newcommand \pDD {0.5}
\newcommand \pE {1.2}
\newcommand \pF {1.9}

\newcommand \kA {k}
\newcommand \kB {l}

\newcommand \doutside{\dalpha+\ddelta+0.4}

\newcommand\xmarker[3]{
 \draw[<->] (#1,\doutside) -- (#2,\doutside);
 \draw (#1,\doutside-0.1) -- ++(0,0.2);
 \draw (#2,\doutside-0.1) -- ++(0,0.2);
 \node at (#1/2 +#2/2 ,\doutside-0.1) {\tiny $#3$};
}
\newcommand\ymarker[3]{
 \draw[<->] (\doutside,#1) -- (\doutside,#2);
 \draw (\doutside-0.1,#1) -- ++(0.2,0);
 \draw (\doutside-0.1,#1) -- ++(0.2,0);
 \draw (\doutside-0.1,#2) -- ++(0.2,0);
 \node[rotate=90] at (\doutside-0.1,#1/2 +#2/2) {\tiny $#3$};
}

\tikzstyle{terminal}=[draw,fill=black, inner sep=1.5pt]

\tikzstyle dterminal=[terminal,pin={[pin 
distance=-1mm,terminal]135:{}}]

\tikzstyle{Steinerpoint}=[circle,draw,fill=white, inner sep=1pt]

\tikzstyle dSteinerpoint=[Steinerpoint,pin={[pin 
distance=-1mm,Steinerpoint]135:{}}]

\tikzstyle{cascade}=[draw,scale=0.55,fill=black,rotate=45]
\tikzstyle{whitevertex}=[draw,scale=0.4,fill=white]
\tikzstyle{Steinerpointcascade}=[draw,fill=white, inner sep=1pt,rotate=45]

\newcommand\OutputI[1]{
  \node[terminal,label=above:{#1}] (t1) at (\pA,\pD) {};
  \draw[thick,->] (\pA,\pC) -- ++(-0.2,0);
  \draw[thick,->] (\pA,\pD) -- ++(-0.2,0);

  \draw (\pA,0.05) ellipse (0.1cm and 0.3cm);
}

\newcommand\OutputII[1]{
  \node[terminal,label=above:{#1}] (t1) at (\pD,\pA) {};
  \draw[thick,->] (\pC,\pA) -- ++(0,-0.2);
  \draw[thick,->] (\pD,\pA) -- ++(0,-0.2);

  \draw (0.05, \pA) ellipse (0.3cm and 0.1cm);
}

\newcommand\InputI[1]{
  \draw[thick,<-] (\pF,\pC) -- ++(0.2,0);
  \draw[thick,<-] (\pF,\pD) -- ++(0.2,0);

  \draw (\pF,0.05) ellipse (0.1cm and 0.3cm);
}

\newcommand\InputII[1]{
  \draw[thick,<-] (\pC,\pF) -- ++(0,0.2);
  \draw[thick,<-] (\pD,\pF) -- ++(0,0.2);

\draw (0.05, \pF) ellipse (0.3cm and 0.1cm);

}

\title{The Depth-Restricted Rectilinear Steiner Arborescence Problem is NP-complete}
\author{Jens Ma{\ss}berg
\footnote{Institut f{\"u}r Optimierung und Operations 
Research, University of Ulm, jens.massberg@uni-ulm.de
}}

\begin{document}

\maketitle



\begin{abstract}

In the rectilinear Steiner arborescence problem the task is to build a shortest
rectilinear Steiner tree connecting a given root and a set of terminals
which are placed in the plane such that all root-terminal-paths are shortest
paths. This problem is known to be NP-hard.

In this paper we consider a more restricted version of this problem. 
In our case we have a depth restrictions  $d(t)\in\mathbb{N}$ for
every terminal $t$. We are looking for a shortest binary rectilinear Steiner
arborescence such that each terminal $t$ is at depth $d(t)$, that is, there are 
exactly $d(t)$ Steiner points on the unique root-$t$-path is exactly $d(t)$.
We prove that even this restricted version is NP-hard.
\end{abstract}

\textbf{keyword:}
Steiner arborescence,
depth restrictions,
NP-completeness,
VLSI design,
shallow light Steiner trees


\section{Introduction}

\subsection{Problem Description}

Let $T=\{t_1,\ldots,t_n\}$ be a set of terminals with positions
$p(t)\in\mathbb{R}^2$ in the plane for all $t\in T$, a 
distinguished terminal $r=t_1$, which we call the \emph{root}, with $p(r)=(0,0)$
and a function $d:T\setminus\{r\}\rightarrow \mathbb{N}$.
A \emph{Depth-Restricted Rectilinear Steiner Arborescence}
is an arborescence $A$ with root $r$, leaves $T\setminus\{r\}$ 
and an embedding $\pi:T\rightarrow \mathbb{R}^2$ in the plane such that
\begin{itemize}
 \item each Steiner point, that is, each vertex of $A$ that is not a terminal, 
   has degree $3$,
 \item each terminal $t\in T$ has degree $1$,
 \item $\pi(t)=p(t)$ for all $t\in T$,
 \item the unique path $P$ in $A$ from $r$ to $t\in T$ is a shortest path with 
 respect to rectilinear distances, that is,
$$\sum_{(v,w)\in E(P)} ||\pi(v)-\pi(w)||_1 = ||p(t)-p(s)||_1$$
 and
 \item for each $t\in T$ the number of internal vertices on the unique path from 
$r$ to $t$ is $d(t)$.
\end{itemize}
The task is to compute such a tree of minimum rectilinear length.
During this paper the depth of a terminal $t$ is always the number of internal 
vertices on the unique $r$-$t$-path.

Note that vertices of a feasible tree might be placed on the same position.
Moreover, in an optimal solution it is possible to have edges that cross or
run parallel on top of each other, which is not possible in an optimal Steiner 
arborescence without depth-restrictions.

%
%

\subsection{Motivation and previous work}

The problem is motivated by an application in VLSI design. 
In the repeater tree problem a signal has to be distributed from a source/root 
to several sinks placed on a chip  by a tree-like network $A$ consisting of 
vertical and horizontal wires. 
The signal is delayed on its way from the source $s$ to each sink $t$, where the
delay can be approximately measured as the length of the unique $s$-$t$-path in 
$A$ plus a constant times the number of internal vertices of the path.
Moreover, for every sink the signal has to arrive before a given individual
arrival time. If we choose these arrival times as small as possible such that
a feasible repeater tree still exists, this problem is equivalent to the 
problem studied in this paper: For each sink
$t$, the source-$t$-path must be a shortest path and the number of
vertices on this path is given by the difference of its length and the arrival
times. See \cite{bartoschek2010repeater} for further details.

The minimum depth-restricted rectilinear Steiner arborescence problem is closely 
related to Steiner trees and Steiner arborescences.
Hwang~\cite{hwang} proved that the classical rectilinear Steiner tree problem is
NP-hard. 
In the rectilinear Steiner arborescence problem the task is to compute a 
rectilinear Steiner tree for a given set of terminals and a distinguished root such that
all root-terminal-paths are shortest paths.
Computing a shortest minimum rectilinear Steiner arborescence is NP-complete
(see Shi and Su~\cite{ShiSu}).

Our problem is an even more restricted version of Stei\-ner arborescences. 
We do not only require that all root-terminal-paths are shortest ones, but 
additionally the depth of each treminal is given. We prove that 
even this restricted version of the problem is NP-complete.

\begin{figure}[ht]
\centering
\begin{tikzpicture}[scale=0.5] 
  \node[terminal,label=left:$r$] at (0,1) {};
  \node[terminal] at (1,4) {};
  \node[terminal] at (4,2) {};
  \node[terminal] at (5,0) {};
  \node[terminal] at (5,3) {};
  \draw (0,1) -- (1,1) -- (1,4);
  \draw (1,2) -- (5,2);
  \draw (5,0) -- (5,3);
  \node[Steinerpoint] at (1,2) {};  
  \node[Steinerpoint] at (5,2) {};  

  \node at (2.7,-0.5) {(i)};
  
  \begin{scope}[xshift=7.5cm]
    \node[terminal,label=left:$r$] at (0,1) {};
    \node[terminal] at (1,4) {};
    \node[terminal] at (4,2) {};
    \node[terminal] at (5,0) {};
    \node[terminal] at (5,3) {}; 
    \draw (5,0) -- (5,3);
    \draw (0,1) -- (5,1);
    \draw (1,1) -- (1,4);
    \draw (4,1) -- (4,2);
    \node at (2.7,-0.5) {(ii)};
    \node[Steinerpoint] at (1,1) {};  
    \node[Steinerpoint] at (4,1) {};  
    \node[Steinerpoint] at (5,1) {};  
  \end{scope}

  \begin{scope}[xshift=15cm]
    \node[terminal,label=left:$r$] at (0,1) {};
    \node[terminal,label=above:$2$] at (1,4) {};
    \node[terminal,label=above:$2$] at (4,2) {};
    \node[terminal,label=right:$2$] at (5,0) {};
    \node[terminal,label=right:$2$] at (5,3) {};
    \draw (5,0) -- (5,3);
    \draw (0,1) -- (5,1);
    \draw (1,1) -- (1,4);
    \draw (1,2) -- (4,2);
    \node at (2.7,-0.5) {(iii)};
    \node[Steinerpoint] at (1,1) {};  
    \node[Steinerpoint] at (1,2) {};  
    \node[Steinerpoint] at (5,1) {};  
  \end{scope}

\end{tikzpicture}
  \caption{A shortest rectilinear Steiner tree (i), a shortest rectilinear 
   Steiner arborescence (ii) and a shortest depth-restricted rectilinear 
   Steiner arborescence (iii). The root of the instances is denoted by $r$ and 
   the numbers in (iii) denote the given depths of the terminals.}
\label{fig:trees}
\end{figure}
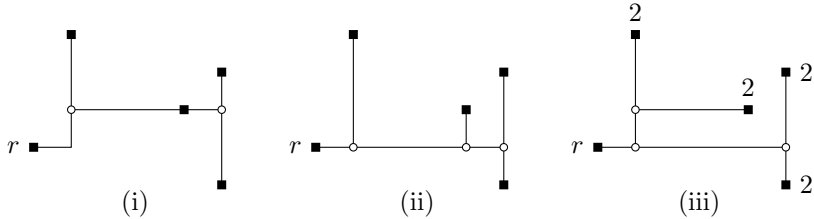

Figure \ref{fig:trees} shows examples for minimum Steiner trees, Steiner arborescences and
depth-restricted Steiner arborescences in the rectilinear plane.

Our trees can be interpreted as shallow-light Steiner arborescences
with vertex delays \cite{heldrotter}. Our results imply that
computing such arborescences in the rectilinear plane is NP-hard.

\section{Feasibility}
It is easy to verify, whether there exists a feasible solution for a given 
instance $(T,p,d)$.
To this end, let $(A,\pi)$ be a feasible solution, that is, $A$ is a binary 
tree 
where each terminal $t\in T$ is at depth $d(t)$.

By Kraft's inequality \cite{kraft}, there exists a binary tree with leaves at
depths exactly $d_1,\ldots, d_n$ (in any order) if and only if 
\begin{equation}\label{kraft}
\sum_{i=1}^n 2^{-d_i} =1.
\end{equation}
Hence, for a given instance $(T,p,d)$ a feasible solution exists if and only  
if (\ref{kraft})
is satisfied for $d_i=d(v_i)$, $i\in\{1,\ldots, n\}$. 
In this case, it is easy to construct a feasible - but not necessarily shortest 
-
depth-restricted Steiner arborescence: Using Huffman coding
\cite{huffman1952method} one can compute a binary tree satisfying (\ref{kraft}).
Placing all internal vertices of this tree
at the position of the root results in a feasible tree.
We conclude that deciding whether a feasible tree exists for a given instance 
can be done in polynomial time. However, we are interested in the complexity of 
computing a shortest depth restricted tree.

\section{Main Idea}

In the remainder of this paper we only consider instances where the root is 
placed at the origin and all terminals are 
placed in the first quadrant.
Note that this is a further restriction of the problem. It will simplify our 
analysis significantly.
Thus in any feasible solution the parent $w$ of a 
vertex $v$ always satisfies $p_x(w)\leq p_x(v)$ and $p_y(w)\leq p_y(v)$.

Let $A$ be a feasible arborescence for an instance $(T,p,d)$. 
Then obviously the depth of an internal vertex $v$ is one smaller than the 
depth of its two children $w_1$ and $w_2$, that is $d(v)=d(w_1)-1 (=d(w_2)-1)$.
We extend the definition of depth to edges by setting the depth of an edge 
$(v,w)$ to be $d(w)$. Two vertices can have a common parent if and only if they 
have to be at the same depth.

If the arborescence $A$ is given, the optimal positions of the internal vertices 
can be easily computed:
\begin{proposition}\label{prop:opt_pos}
 If an internal vertex $v$ has two children at positions $(x_1, y_1)$ and $(x_2, 
 y_2)$, respectively, then the optimal position for $v$ is $(\min\{x_1, 
 x_2\},$ $\min\{y_1,$ $y_2\})$.
\end{proposition}
{\it Proof:}
 In a feasible solution we have $p_x(v) \leq \min\{x_1,$ $x_2\}$
 and $p_y(v) \leq \min\{y_1,$ $y_2\}$. If one of the inequalities is not 
satisfied 
 with equality, moving the vertex to the right or above yields another feasible 
 embedding that is shorter.
\endproof

An intermediate consequence of Proposition \ref{prop:opt_pos} is that in every 
optimal 
solution all vertices are placed on the vertices of the Hanan grid on $S$; that 
is, for every internal vertex at position $(x,y)$ there exist two terminals 
with x- and y-coordinate $x$ and $y$, respectively (see 
\cite{hanan1966steiner}).

\section{Reduction}

\subsection{Reduction Overview}

We prove the NP-completeness by a reduction from \emph{Maximum 
2-Satis\-fi\-ability} 
(in short \emph{Max-2-Sat}). A Max-2-Sat instance consists of a set of 
variables
$\mathcal{V}=\{x_1,\ldots, x_n\}$ and a set of clauses 
$\mathcal{C}=\{C_1,\ldots, C_m\}$ on $\mathcal{V}$ with $|C_i|=2$ for 
$i\in\{1,\ldots, 
m\}$.
The problem is to find a truth assignment $\pi$ such that the number of clauses 
 satisfied by $\pi$ is maximized.
Garey et al.~\cite{gareyjohnsonstockmeyer} proved that Max-2-Sat is NP-hard.

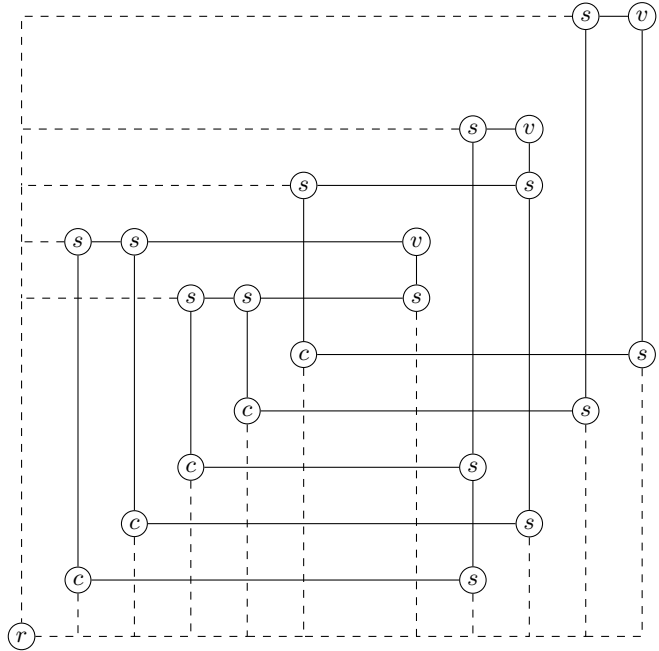
\begin{figure}[ht]
\begin{center}
 \begin{tikzpicture}[scale=0.75]
  \node[draw, circle, inner sep=1.5pt] (r) at (0,0) {\small $r$};
  \node[draw, circle, inner sep=1.5pt] (c1) at (1,1) {\small $c$};
  \node[draw, circle, inner sep=1.5pt] (c2) at (2,2) {\small $c$};
  \node[draw, circle, inner sep=1.5pt] (c3) at (3,3) {\small $c$};
  \node[draw, circle, inner sep=1.5pt] (c4) at (4,4) {\small $c$};
  \node[draw, circle, inner sep=1.5pt] (c5) at (5,5) {\small $c$};

\node[draw, circle, inner sep=1.5pt] (x1) at (7,7) {\small $v$};
\node[draw, circle, inner sep=1.5pt] (x2) at (9,9) {\small $v$};
\node[draw, circle, inner sep=1.5pt] (x3) at (11,11) {\small $v$};

\node[draw, circle, inner sep=1.5pt] (s1) at (1,7) {\small $s$};
\node[draw, circle, inner sep=1.5pt] (s2) at (2,7) {\small $s$};
\node[draw, circle, inner sep=1.5pt] (s3) at (3,6) {\small $s$};
\node[draw, circle, inner sep=1.5pt] (s4) at (4,6) {\small $s$};
\node[draw, circle, inner sep=1.5pt] (s5) at (5,8) {\small $s$};
\node[draw, circle, inner sep=1.5pt] (s7) at (8,9) {\small $s$};
\node[draw, circle, inner sep=1.5pt] (s8) at (10,11) {\small $s$};

\node[draw, circle, inner sep=1.5pt] (s10) at (8,1) {\small $s$};
\node[draw, circle, inner sep=1.5pt] (s11) at (9,2) {\small $s$};
\node[draw, circle, inner sep=1.5pt] (s12) at (8,3) {\small $s$};
\node[draw, circle, inner sep=1.5pt] (s13) at (10,4) {\small $s$};
\node[draw, circle, inner sep=1.5pt] (s14) at (11,5) {\small $s$};
\node[draw, circle, inner sep=1.5pt] (s15) at (7,6) {\small $s$};
\node[draw, circle, inner sep=1.5pt] (s16) at (9,8) {\small $s$};

\draw (c1) -- (s1) -- (s2) -- (c2);
\draw (s2) -- (x1);
\draw (c3) -- (s3) -- (s4) -- (c4);
\draw (s4) -- (s15) -- (x1);
\draw (c5) -- (s5)  -- (s16) -- (x2);
\draw (c1) -- (s10) -- (s12) -- (c3);
\draw (s12) -- (s7) -- (x2);
\draw (c2) -- (s11) -- (s16);
\draw (c4) -- (s13) -- (s8) -- (x3);
\draw (c5) -- (s14) -- (x3);

\draw[dashed] (c1) -- (1,0) -- (r);
\draw[dashed] (c2) -- (2,0) -- (1,0);
\draw[dashed] (c3) -- (3,0) -- (2,0);
\draw[dashed] (c4) -- (4,0) -- (3,0);
\draw[dashed] (c5) -- (5,0) -- (4,0);
\draw[dashed] (s15) -- (7,0) -- (5,0);
\draw[dashed] (s10) -- (8,0) -- (7,0);
\draw[dashed] (s11) -- (9,0) -- (8,0);
\draw[dashed] (s13) -- (10,0) -- (9,0);
\draw[dashed] (s14) -- (11,0) -- (10,0);

\draw[dashed] (s3) -- (0,6) -- (r);
\draw[dashed] (s1) -- (0,7) -- (0,6);
\draw[dashed] (s5) -- (0,8) -- (0,7);
\draw[dashed] (s7) -- (0,9) -- (0,8);
\draw[dashed] (s8) -- (0,11) -- (0,9);

 \end{tikzpicture}
 \caption{High-level overview of the transformed instance. 
 Root, variable, clause and splitter gadgets are marked by
 r,v,c and s, respectively. 
The bold lines show the connections between the gadgets and the dashed lines
the connections required to enable a feasible solution. }
 \label{fig:high-level}
\end{center}
\end{figure}

We use the component design technique to transform a Max-2-Sat instance
$(\mathcal{V},\mathcal{C})$ into an 
instance for our problem.
First we give a high-level overview of this reduction.
The construction consists of several types of gadgets that are placed on a 
uniform grid, where the root is located at the origin.
For each variable and each clause, we have a variable gadget and clause gadget,
respectively, that are placed on the diagonal of the grid (see Figure 
\ref{fig:high-level}). 
The gadgets are connected by horizontal or vertical connections representing 
the 
literals: For every variable $x_i$, one connection is leaving the variable 
gadget to the left (corresponding to the literal $x_i$) and one connection is 
leaving to the bottom (corresponding to $\overline{x_i}$).
Each clause gadget receives one connection from above and one from the right, 
representing the corresponding literals of the clause.
In order to split a connection and to switch from horizontal to vertical
connections or vice versa
 we add splitter gadgets. Finally, we require connections ensuring the 
existence 
of a feasible solution for the instance (marked by dashed lines in Figure 
\ref{fig:high-level}).

Each truth assignment for $(\mathcal{V},\mathcal{C})$ corresponds to a feasible 
Steiner arborescence where a
truth assignment satisfying a maximal number of clauses corresponds to a 
shortest feasible Steiner arborescence.
The length of the connections within the variable and splitter gadgets 
and between them differ only slightly for different truth assignments, but the 
length of a connection of clause gadgets increases by a (relatively large) 
constant $C$ if the clause is not satisfied by the truth assignment.
Therefore, there exists a truth assignment $\pi$ satisfying $k$ of $m$ clauses 
if and only if there exists a feasible Steiner arborescence of length at most
$c+(m-k)C$ where $c$ is a constant that is independent of $\pi$.

\subsection{Tile design}
The gadgets are realized by equal sized quadratic \emph{tiles}.
In this section we describe the design of the different types of tiles.
Each tile has size $(4\alpha+2)\times (4\alpha+2)$ and contains several 
terminals, depending on the tile's type. 
The tiles are placed on a uniform grid containing $(1+m+2n)\times
(1+m+2n)$ tiles and having lattice spacing $4\alpha+2$. The integral constant $\alpha$ 
 will be set later (see  Section 
\ref{section:values}). 

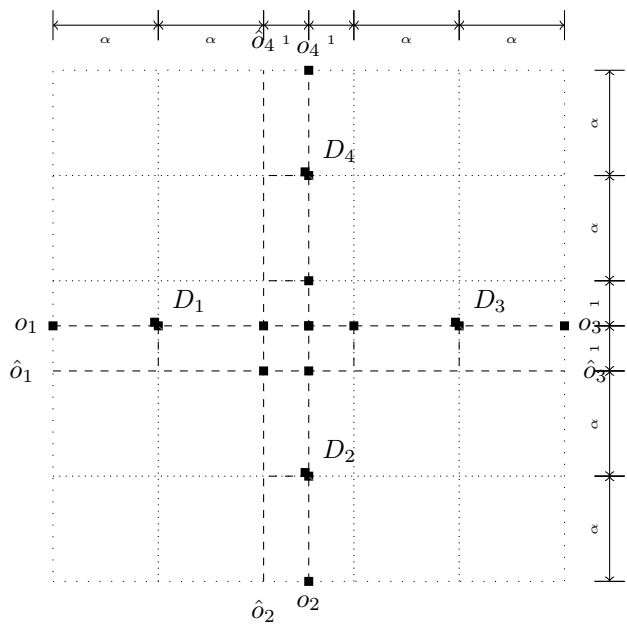
\begin{figure}

\begin{tikzpicture}[scale=2]
  \tileborder

  \xmarker{\pA}{\pB}{\alpha};
  \xmarker{\pB}{\pC}{\alpha};
  \xmarker{\pC}{\pD}{1};
  \xmarker{\pD}{\pDD}{1};
  \xmarker{\pDD}{\pE}{\alpha};
  \xmarker{\pE}{\pF}{\alpha};

  \ymarker{\pA}{\pB}{\alpha};
  \ymarker{\pB}{\pC}{\alpha};
  \ymarker{\pC}{\pD}{1};
  \ymarker{\pD}{\pDD}{1};
  \ymarker{\pDD}{\pE}{\alpha};
  \ymarker{\pE}{\pF}{\alpha};

  \node[terminal, label=above:{$o_4$}] at (\pD,\pF) {};
  \node[label=above:{$\hat{o}_4$}] at (\pC,\pF) {};
  \node[terminal, label=right:{$o_3$}] at (\pF,\pD) {};
  \node[label=right:{$\hat{o}_3$}] at (\pF,\pC) {};

  \node[terminal, label=left:{$o_1$}] at (\pA,\pD) {};
  \node[label=left:{$\hat{o}_1$}] at (\pA,\pC) {};
  \node[dterminal, label=above right:{$D_1$}] at (\pB,\pD) {};
  \node[terminal] at (\pC,\pD) {};
  \node[terminal] at (\pD,\pD) {};
  \node[dterminal, label=above right:{$D_3$}] at (\pE,\pD) {};

  \node[terminal, label=below:{$o_2$}] at (\pD,\pA) {};
  \node[label=below:{$\hat{o}_2$}] at (\pC,\pA) {};
  \node[dterminal, label=above right:{$D_2$}] at (\pD,\pB) {};
  \node[terminal] at (\pD,\pC) {};
  \node[terminal] at (\pD,\pD) {};
  \node[dterminal, label=above right:{$D_4$}] at (\pD,\pE) {};

  \node[terminal] at (\pC,\pC) {};

  \node[terminal] at (\pD,\pDD) {};
  \node[terminal] at (\pDD,\pD) {};

  \draw[dotted] (\pA, \pB) -- (\pF, \pB);
  \draw[dotted] (\pA, \pDD) -- (\pF, \pDD);
  \draw[dotted] (\pA, \pE) -- (\pF, \pE);
  \draw[dotted] (\pB, \pA) -- (\pB, \pF);
  \draw[dotted] (\pDD, \pA) -- (\pDD, \pF);
  \draw[dotted] (\pE, \pA) -- (\pE, \pF);

  \draw[dashed] (\pA, \pD) -- (\pF, \pD);
  \draw[dashed] (\pA, \pC) -- (\pF, \pC);
  \draw[dashed] (\pD, \pA) -- (\pD, \pF);
  \draw[dashed] (\pC, \pA) -- (\pC, \pF);
	
  \draw[dashed] (\pB, \pD) -- (\pB, \pC);
  \draw[dashed] (\pDD, \pD) -- (\pDD, \pC);
  \draw[dashed] (\pE, \pD) -- (\pE, \pC);
  \draw[dashed] (\pD, \pB ) -- (\pC, \pB);
  \draw[dashed] (\pD, \pDD) -- (\pC, \pDD);
  \draw[dashed] (\pD, \pE ) -- (\pC, \pE);

\end{tikzpicture}
\caption{Prototype of a tile.}
\label{fig:prototype}

\end{figure}

Figure \ref{fig:prototype} shows a prototype of a tile.
The black squares show possible positions of terminals and the dotted lines 
show the Hanan grid on the set of possible terminal positions. 

On some type of tiles we use \emph{terminal cascades}, a set of terminals
with consecutive given depths placed at the same position.
If an edge of depth $a+k$ starts at the position $p$ of a terminal cascade
containing $k$ terminals with depths from $a+1$
to $a+k$, then  all terminals of the cascade can be connected to the tree by
adding $k$ Steiner points at position $p$ and connecting the terminals and 
Steiner points appropriately.
In this case, an edge of depth $a$ ends at the cascade and we have no 
additional connection cost as all inserted edges have length $0$. 
If no edge of depth $a+k$ starts at the position of the cascade, 
the instance is constructed such that
we have $k$ edges of length at least $1$, increasing the cost to connect the 
cascade to the tree by at least $k$.

A special type of terminal cascades are the \emph{double terminals} consisting
of two terminals at the same position with consecutive depths. Double 
terminals are only placed on positions $D_j$, $j\in\{1,2,3,4\}$ (see Figure 
\ref{fig:prototype}).
A tile $t$ contains a terminal at position $o_j$ (for $j\in\{1,2,3,4\}$) if and 
only if there is a double terminal at position $D_j$.
Moreover, if the double terminal at $D_j$, $j\in\{1,2\}$, has depths $k$ 
and $k-1$, then $o_j$ has 
depth $k-2$ and if the double terminal at $D_j$, $j\in\{3,4\}$ has
depths $k$ and $k-1$, then $o_j$ has depth $k+1$.

Let $(A,\pi)$ be a feasible solution, $t$ a tile and $\pi|_t$ be the embedding 
we obtain by projecting all vertices that are outside of $t$ with respect to 
$\pi$ to the nearest point on the border of $t$.
Then the \emph{length of $(A,\pi)$ on $t$} is the total length with respect to 
$\pi|_t$ of all edges $(v,w)$ that contain at least one vertex in the inner of 
$t$.
Note that the total length of $(A,\pi)$ on all tiles is a lower bound for the length 
of $(A,\pi)$.

\begin{proposition}\label{prop:doubleterminals}
 If $t$ is a tile with $k$ double terminals, then every feasible 
solution has length at least $2k\alpha$ on $t$.
\end{proposition}
{\it Proof:} 
 Recall that double terminals are only located at positions $D_j$, 
 $j\in\{1,2,3,4\}$.
 Let $H$ be the Hanan grid on all terminals and consider 
 a feasible arborescence $A$ so that all vertices are on positions of 
 vertices of $H$.
 Let $D$ be a double terminal at position $p$ and $P$ be the set of vertices of 
 $A$ having distance at most 1 to $p$.
 By construction, for every edge $\{v,w\}$ with $v\in P$ we have either $w\in 
P$ 
 or $||p(v)-p(w)||_1\geq \alpha$.

 In a feasible arborescence at least one edge must leave $P$. If no edge is 
 entering $P$, then it only contains the two terminals at positions $p$. As 
 they have different depths, the cannot have the same parent and thus two 
 edges must leave $P$. Thus $|\delta(P)|\geq 2$  implies that connecting a 
 double terminal increases the length of a feasible  connection by at least 
 $2\alpha$.
\endproof 

An intermediate consequence is the following:
\begin{conclusion}
 Any feasible arborescence for an instance containing $k$ double terminals has 
length at least $2k\alpha$.
\end{conclusion}

We construct the tiles and the instance $I$ such that if $I$ contains 
$k$ double terminals, then there exists a solution $A$ of length less than
$(2k+1)\alpha$.

\begin{lemma}\label{lemma:puts}
 Let $(A,\pi)$ be an optimal solution of length strictly less than $(2k+1)\alpha$ for 
an 
 instance $I$ with $k$ double terminals.
 If $t$ is a tile with terminal $s$ at position $o_i$ for some 
$i\in\{1,2,3,4\}$, then 
 the Steiner point connected to $s$ is either placed at position $o_i$ or 
 position $\hat{o}_i$.
\end{lemma}
{\it Proof:} 
 Denote by $s'$ the Steiner point in $A$ connected to $s$. As $A$ is optimal,
 $s'$ is placed on the Hanan grid given by the terminals. If $s'$ is not placed 
 at $o_i$ or $\hat{o}_i$, then the distance between $s$ and $s'$ is at 
 least $\alpha$. But then the total length of $A$ is at least $(2k+1)\alpha$, 
 contradicting the assumption.
\endproof 

During the remainder of the paper we make the following assumption on the 
constructed instances. By setting $\alpha$ to an appropriate value, 
we can later guarantee that the assumption is indeed satisfied.
\begin{assumption}\label{ass1}
  If $I$ is an instance with $k$ double terminals, then there exists an optimal 
solution of cost less than $(2k+1)\alpha$.
\end{assumption}

By Lemma \ref{lemma:puts} a tile $t$ can only be entered or left at
Steiner points that are connected to a terminal at position $o_i$, 
$i\in\{1,2,3,4\}$.
We call these Steiner points \emph{input} or \emph{output} of tile $t$ if it is 
connected to a terminal at position $o_i$ with $i\in \{3,4\}$ or $i\in\{1,2\}$, 
respectively. Note that each input of a tile $t$ is the output of the tile that 
shares the border of $t$ containing the input.
Thus it suffices to only consider inputs in the following.
For every input $s$, we define the depth and the  \emph{parity} of $s$:
Consider a tile with an input connected to a terminal $v$ at position $o_i$.
The depth of $s$ is $d(v)-1$ (recall that $s$ is a 
child of $v$).
If the input is placed at the position of $v$, then the parity of the 
input is $0$. Otherwise, the parity is $1$.
Later we associate with each input a literal. Then the literal is set to 
$\true$ if and only if the parities of the associated inputs are $1$.

Now we can decompose an optimal solutions at its inputs and outputs:
Consider a tile $t$ with inputs $I$ and parities $\pi:I\rightarrow \{0,1\}$ 
for the inputs. A \emph{tile branching} for $(t,\pi)$ is a branching $B$
(that is a forest where each tree is an arborescence)
containing an arborescence for each output and the leaves of $B$ are the 
terminals of $t$ plus one leaf for each input $i\in I$ at the position 
corresponding to the parity $\pi(i)$. Moreover, the branching satisfies the 
depth-restrictions, that is, $d(w)=d(v)-1$ for the parent $w$ of a vertex 
$v$. Hereby we use the depths of the inputs as the depths for the 
terminals placed at their positions. An implication is that the depths of 
the roots coincide with the depths of the corresponding outputs.

A \emph{shortest connection} for tile $t$ with parities $\pi$ for the inputs is 
a shortest tile branching for $(t,\pi)$.
Shortest connections for a given tile $t$ with parities $\pi$ can be computed 
efficiently by dynamic programming: We know that the Steiner points are only 
placed at vertices of the Hanan grid and that both children of a Steiner point 
must have the same depth.

\begin{figure}
\begin{tikzpicture}[scale=1.3]
  \node[terminal] at (0.2,0) {};
  \node at (1.5,0) {terminal};
  \node[cascade] at (0.2,-0.5) {};
  \node at (1.5,-0.5) {terminal cascade};
  \node[dterminal] at (0.2,-1) {};
  \node at (1.5,-1) {double terminal};

  \node[Steinerpoint] at (3.0,0) {};
  \node at (4.5,0) {Steiner point};
  \node[Steinerpointcascade] at (3.0,-0.5) {};
  \node at (4.5,-0.5) {Steiner cascade};

  \node[terminal] at (3.0,-1) {};
  \node[Steinerpoint] at (3.0,-1) {};
  \node at (4.6,-1) {Steiner point at terminal};

\end{tikzpicture}
\label{fig:legend}
\caption{Explanations.}
\end{figure}

\subsection{Variable Tiles}

\begin{figure}
\begin{center}
 \begin{tikzpicture}[scale=1.5]
  \tileborder

  \xmarker{\pA}{\pB}{\alpha};
  \xmarker{\pB}{\pC}{\alpha};
  \xmarker{\pC}{\pD}{1};
  \xmarker{\pD}{\pF}{2\alpha +1};

  \ymarker{\pA}{\pB}{\alpha};
  \ymarker{\pB}{\pC}{\alpha};
  \ymarker{\pC}{\pD}{1};
  \ymarker{\pD}{\pF}{2\alpha +1};

  \OutputI{$k$-$3$}

  \OutputII{$k$-$3$}

  \node[terminal,label=above right:{$k$}] (a1) at (\pD,\pD) {};
  \node[terminal,label=above:{$k$}] (a2) at (\pC,\pD) {};
  \node[terminal,label=right:{$k$}] (a3) at (\pD,\pC) {};
  \node[terminal,label=below left:{$k$}] (a4) at (\pC,\pC) {};

  \node[dterminal,label=above:{$k$-$1,k$-$2$}] (b12) at (\pB,\pD) {};
  \node[dterminal,label=right:{$k$-$1,k$-$2$}] (b22) at (\pD,\pB) {};

  \draw[dotted,thick] (a1) -- (\pA,\pD);
  \draw[dotted,thick] (a3) -- (a4) -- (\pC,\pA);
  \draw[dotted,thick] (a1) -- (\pD,\pA);
  \draw[dotted,thick] (a2) -- (a4) -- (\pA,\pC);
  \draw[dotted,thick] (b12) -- (\pB,\pC);
  \draw[dotted,thick] (b22) -- (\pC,\pB);

 \end{tikzpicture}
 \caption{A variable tile}
 \label{fig:variabletile}
\end{center}
\end{figure}
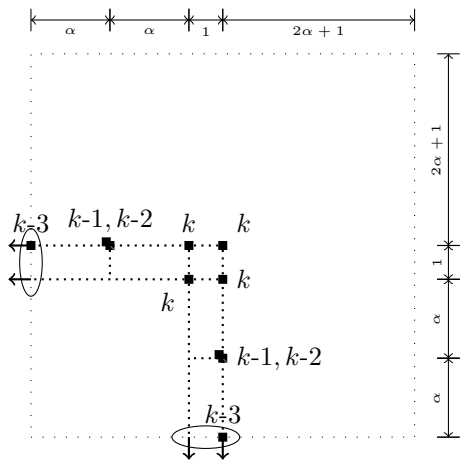

For every variable of a 3-SAT instance $(\mathcal{V},\mathcal{C})$ we build a 
\emph{variable tile} containing 8 terminals  
and two outputs. 
Figure \ref{fig:variabletile} shows the positions of the terminals as black 
squares and their depths. The dotted lines show possible positions of edges 
in an optimal solution.

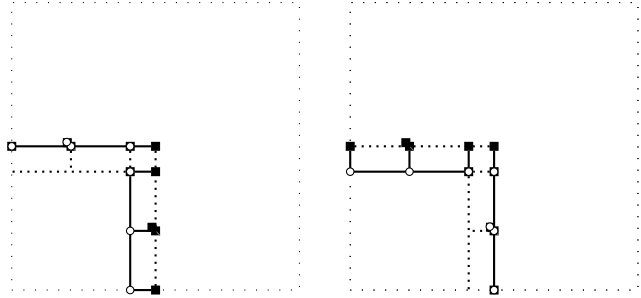
\begin{figure}
\begin{center}
 \begin{tikzpicture}[scale=1.5]

   \begin{scope}[xshift=0cm,yshift=3cm, scale=0.75]
  \begin{scope}
   \tileborder
   \node[terminal] (a1) at (\pD,\pD) {};
   \node[terminal] (a2) at (\pC,\pD) {};
   \node[terminal] (a3) at (\pD,\pC) {};
   \node[terminal] (a4) at (\pC,\pC) {};

   \node[dterminal] (b12) at (\pB,\pD) {};
   \node[dterminal] (b22) at (\pD,\pB) {};

   \node[terminal] (o1) at (\pA,\pD) {};
   \node[terminal] (o2) at (\pD,\pA) {};

   \draw[dotted,thick] (a1) -- (\pD,\pA);
   \draw[dotted,thick] (a2) -- (a4) -- (\pA,\pC);
   \draw[dotted,thick] (b12) -- ++(0,-\ddelta);
   \draw[thick] (b22) -- ++(-\ddelta,0);

   \draw[thick] (a1) -- (\pA,\pD);
   \draw[thick] (a3) -- (a4) -- (\pC,\pA);
   \draw[thick] (o2) -- (\pC,\pA);
  
   \node[Steinerpoint] at (\pC,\pD) {};
   \node[Steinerpoint] at (\pB,\pD) {};
   \node[Steinerpoint] at (\pB-0.05,\pD+0.05) {};

   \node[Steinerpoint] at (\pC,\pC) {};
   \node[Steinerpoint] at (\pC,\pB) {};

   \node[Steinerpoint] at (\pA,\pD) {};
   \node[Steinerpoint] at (\pC,\pA) {};
  \end{scope}

  \begin{scope}[xshift=4cm, yshift=0cm]
   \tileborder
   \tileborder
   \node[terminal] (a1) at (\pD,\pD) {};
   \node[terminal] (a2) at (\pC,\pD) {};
   \node[terminal] (a3) at (\pD,\pC) {};
   \node[terminal] (a4) at (\pC,\pC) {};

   \node[dterminal] (b12) at (\pB,\pD) {};
   \node[dterminal] (b22) at (\pD,\pB) {};

   \node[terminal] (o1) at (\pA,\pD) {};
   \node[terminal] (o2) at (\pD,\pA) {};

   \draw[thick] (a1) -- (\pD,\pA);
   \draw[thick] (a2) -- (a4) -- (\pA,\pC);
   \draw[thick] (b12) -- ++(0,-\ddelta);

   \draw[dotted,thick] (b22) -- ++(-\ddelta,0);
   \draw[dotted,thick] (a1) -- (\pA,\pD);
   \draw[dotted,thick] (a3) -- (a4) -- (\pC,\pA);    

   \draw[thick] (o1) -- (\pA,\pC);

   \node[Steinerpoint] at (\pC,\pC) {};
   \node[Steinerpoint] at (\pB,\pC) {};
   \node[Steinerpoint] at (\pD,\pC) {};
   \node[Steinerpoint] at (\pD,\pB) {};
   \node[Steinerpoint] at (\pD-0.05,\pB+0.05) {};

   \node[Steinerpoint] at (\pA,\pC) {};
   \node[Steinerpoint] at (\pD,\pA) {};
	\end{scope}
 \end{scope}

 \end{tikzpicture}
 \caption{The two possible shortest connections for a variable tile}
 \label{fig:variabletile2}
\end{center}
\end{figure}

\begin{lemma}
 There are two shortest connections for a variable tile. 
 The total edge length of a minimum connection is $4\alpha+5$.
\end{lemma}
{\it Proof:} 
 The two minimum connections are shown in Figure \ref{fig:variabletile2}. 
 Note that by the placement of the terminal at position 
 $(2\alpha+1,2\alpha+1)$ the parity of at least one of the outputs must be $0$.
\endproof 

We associate with the output at position $o_1/\hat{o}_1$  the literal $x_i$ and 
with the output 
at position $o_2/\hat{o}_2$ literal $\overline{x_i}$. Thereby, the 
connection in Figure \ref{fig:variabletile2} (left) corresponds to $x_i=\true$, 
while the connection in Figure \ref{fig:variabletile2} (right) corresponds to 
$x_i=\false$.

\subsection{Clause Tiles}

For every clause, we construct a clause tile as shown in Figure
\ref{fig:clausetile}. 
It contains  two terminal cascades (drawn as black rhombs), each 
with $\beta$ terminals and both inputs have depth $k$. The integral value 
$\beta$ is constant and the same for 
all clause tiles and will be set later.
We denote the upper right terminal cascade by $S_1$.

\begin{figure}[ht]
\begin{center}
 \begin{tikzpicture}[scale=1.5]

  
    \begin{scope}[xshift=-1cm,yshift=-3.5cm, scale=0.75]
     \begin{scope}[xshift=0cm, yshift=0cm]
   \tileborder

   \node[cascade] (a1) at (\pD,\pD) {};
   \node[cascade] (a2) at (\pC,\pC) {};
   \node[dterminal] (a33) at (\pD,\pB) {};
   \node[dterminal] (b12) at (\pD,\pE) {};

   \node[dterminal] (b22) at (\pE,\pD) {};
   \node[terminal] (o3) at (\pD,\pA) {};

   \draw[dotted] (\pD,\pC) -- (\pD,\pF);
   \draw[dotted] (\pC,\pA) -- (\pC,\pF);
   \draw[dotted] (\pC,\pD) -- (\pF,\pD);
   \draw[dotted] (\pC,\pC) -- (\pF,\pC);
   \draw[dotted] (\pD,\pE) -- (\pC,\pE);
   \draw[dotted] (\pE,\pD) -- (\pE,\pC);

   \draw[thick] (\pD,\pF) -- (a1) -- (\pC,\pD) -- (a2) -- (\pC,\pA);
   \draw[thick] (\pF,\pD) -- (\pE,\pD) -- (\pE,\pC) -- (\pC,\pC);
	 \draw[thick] (a33) -- (\pC,\pB);
	 \draw[thick] (o3) -- (\pC,\pA);

   \node[Steinerpointcascade] (a) at (\pD,\pD) {};
   \node[Steinerpointcascade] at (\pC,\pC) {};
   \node[Steinerpoint] (b12) at (\pD,\pE) {};
   \node[Steinerpoint] at (\pD-0.05,\pE+0.05) {};

   \node[Steinerpoint] (b22) at (\pE,\pD) {};
   \node[Steinerpoint] at (\pE-0.05,\pD+0.05) {};

   \node[Steinerpoint] (a33) at (\pC,\pB) {};
   \node[Steinerpoint] at (\pD,\pF) {};
   \node[Steinerpoint] at (\pF,\pD) {};
   \node[Steinerpoint] at (\pC,\pA) {};

  \end{scope}

  \begin{scope}[xshift=3.8cm, yshift=0cm]
  
    \tileborder
   \node[cascade] (a1) at (\pD,\pD) {};
   \node[cascade] (a2) at (\pC,\pC) {};
   \node[dterminal] (a33) at (\pD,\pB) {};
   \node[dterminal] (b12) at (\pD,\pE) {};
   \node[dterminal] (b22) at (\pE,\pD) {};
   \node[terminal] (o3) at (\pD,\pA) {};

   \draw[dotted] (\pD,\pC) -- (\pD,\pF);
   \draw[dotted] (\pC,\pA) -- (\pC,\pF);
   \draw[dotted] (\pC,\pD) -- (\pD,\pD);
   \draw[dotted] (\pC,\pC) -- (\pF,\pC);
   \draw[dotted] (\pD,\pE) -- (\pC,\pE);
   \draw[dotted] (\pE,\pD) -- (\pE,\pC);

    \draw[thick] (\pC,\pF) -- (\pC,\pA);
    \draw[thick] (\pD,\pE) -- (\pC,\pE);
    \draw[thick] (a2) -- (\pD,\pC) -- (a1) -- (\pF,\pD);
    \draw[thick] (a33) -- (\pC,\pB);
    \draw[thick] (o3) -- (\pC,\pA);

   \node[Steinerpointcascade] (a) at (\pD,\pD) {};
   \node[Steinerpointcascade] at (\pC,\pC) {};
   \node[Steinerpoint] (b12) at (\pC,\pE) {};
   \node[Steinerpoint] (b22) at (\pE,\pD) {};
   \node[Steinerpoint] at (\pE-0.05,\pD+0.05) {};

   \node[Steinerpoint] (a33) at (\pC,\pB) {};

   \node[Steinerpoint] at (\pC,\pF) {};
   \node[Steinerpoint] at (\pF,\pD) {};
   \node[Steinerpoint] at (\pC,\pA) {};
  \end{scope}

  \begin{scope}[xshift=0cm, yshift=-3.8cm]
    \tileborder

   \node[cascade] (a1) at (\pD,\pD) {};
   \node[cascade] (a2) at (\pC,\pC) {};
   \node[dterminal] (a33) at (\pD,\pB) {};
   \node[dterminal] (b12) at (\pD,\pE) {};
   \node[dterminal] (b22) at (\pE,\pD) {};
   \node[terminal] (o3) at (\pD,\pA) {};

   \draw[dotted] (\pD,\pC) -- (\pD,\pF);
   \draw[dotted] (\pC,\pA) -- (\pC,\pF);
   \draw[dotted] (\pC,\pD) -- (\pF,\pD);
   \draw[dotted] (\pC,\pC) -- (\pF,\pC);
   \draw[dotted] (\pD,\pE) -- (\pC,\pE);
   \draw[dotted] (\pE,\pD) -- (\pE,\pC);

    \draw[thick] (\pD,\pF) -- (\pD,\pD); 
    \draw[thick] (a1) -- (\pC,\pD) -- (a2) -- (\pC,\pA);
    \draw[thick] (\pF,\pC) -- (\pC,\pC);
    \draw[thick] (\pE,\pD) -- (\pE,\pC);
    \draw[thick] (a33) -- (\pC,\pB);
    \draw[thick] (o3) -- (\pC,\pA);

   \node[Steinerpointcascade] (a) at (\pD,\pD) {};
   \node[Steinerpointcascade] at (\pC,\pC) {};
   \node[Steinerpoint] (b12) at (\pD,\pE) {};
   \node[Steinerpoint] at (\pD-0.05,\pE+0.05) {};

   \node[Steinerpoint] (b22) at (\pE,\pC) {};
   \node[Steinerpoint] (a33) at (\pC,\pB) {};
	
   \node[Steinerpoint] (i1) at (\pF,\pC) {};
   \node[Steinerpoint] (i2) at (\pD,\pF) {};
   \node[Steinerpoint] at (\pC,\pA) {};
  \end{scope}

 \begin{scope}[xshift=3.8cm, yshift=-3.8cm]
    \tileborder

   \node[cascade] (a1) at (\pD,\pD) {};
   \node[cascade] (a2) at (\pC,\pC) {};
   \node[dterminal] (a33) at (\pD,\pB) {};
   \node[dterminal] (b12) at (\pD,\pE) {};
   \node[dterminal] (b22) at (\pE,\pD) {};
   \node[terminal] (o3) at (\pD,\pA) {};

   \draw[dotted] (\pD,\pC) -- (\pD,\pF);
   \draw[dotted] (\pC,\pA) -- (\pC,\pF);
   \draw[dotted] (\pC,\pD) -- (\pF,\pD);
   \draw[dotted] (\pC,\pC) -- (\pF,\pC);
   \draw[dotted] (\pD,\pE) -- (\pC,\pE);
   \draw[dotted] (\pE,\pD) -- (\pE,\pC);

   \draw[thick] (\pF,\pC) -- (\pC,\pC) -- (\pC,\pF);
   \draw[thick] (\pC,\pA) -- (\pC,\pC); 
   \draw[thick] (a33) -- (\pC,\pB);
   \draw[thick] (o3) -- (\pC,\pA);

   \draw[thick] (\pD,\pD) -- (\pD,\pC);
   \draw[thick] (\pC,\pE) -- (\pD,\pE);
   \draw[thick] (\pE,\pC) -- (\pE,\pD);

   \node[Steinerpointcascade] (a) at (\pD,\pC) {};
	 \node[Steinerpoint] at (\pC,\pE) {};
	 \node[Steinerpoint] at (\pE,\pC) {};

   \node[Steinerpointcascade] at (\pC,\pC) {};
   \node[Steinerpoint] (a33) at (\pC,\pB) {};
   \node[Steinerpoint] at (\pC,\pF) {};
   \node[Steinerpoint] at (\pF,\pC) {};
	 \node[Steinerpoint] at (\pC,\pA) {};

  \end{scope}

    \end{scope}
 \end{tikzpicture}
 \caption{Shortest connections for a clause tile}
 \label{fig:clausetile2}
\end{center}
\end{figure}
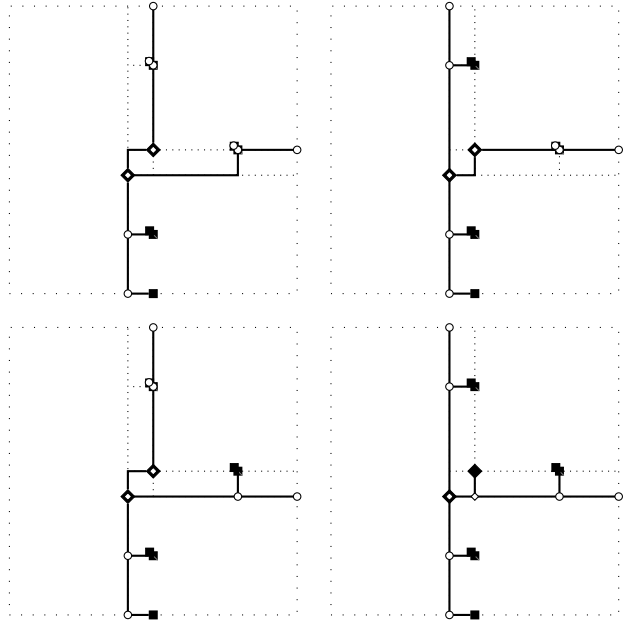

The length of a minimum connection of a clause tile depends on the parities of 
the inputs.
\begin{lemma}\label{lemma:clause}
 A minimum connection for a clause tile has length $6\alpha+9$ if both inputs 
  have parity $1$,  length $6\alpha+10$ if exactly one input has parity $1$ and 
length $6\alpha+11+\beta$ if both inputs have parity $0$.
\end{lemma}
{\it Proof:} 
 Figure \ref{fig:clausetile2} shows minimum connections for the four 
 different pairs of parities of the inputs.
 Note that in the case where both inputs have parity $0$ there are 
 $k-l$ edges of length $1$ leaving the terminal cascade at position 
 $(2\alpha+1,2\alpha+1)$.
\endproof 

The terminal cascade at position 
$(2\alpha,2\alpha)$ enforces the parity of the output to be $0$.

\begin{figure}[ht]
\begin{center}

 \begin{tikzpicture}[scale=1.5]
  \tileborder

  \InputI{$\kA$}
  \InputII{$\kA$}
	\OutputII{$\kA$-$\beta$-$5$}

  \xmarker{\pA}{\pC}{2\alpha};
  \xmarker{\pC}{\pD}{1};
  \xmarker{\pD}{\pE}{\alpha+1};
  \xmarker{\pE}{\pF}{\alpha};
  \ymarker{\pA}{\pB}{\alpha};
  \ymarker{\pB}{\pC}{2\alpha};
  \ymarker{\pC}{\pD}{1};
  \ymarker{\pD}{\pE}{\alpha+1};
  \ymarker{\pE}{\pF}{\alpha};

  \node[cascade,label=above:{$[\kA$-$2,\kA$-$\beta$-$1]$}] (a1) at (\pD,\pD) {};
  \node[cascade,label=left:{$[\kA$-$2,\kA$-$\beta$-$1]$}] (a2) at (\pC,\pC) {};
  \node[dterminal,label=right:{$\kA$-$\beta$-$3$, $\kA$-$\beta$-$4$}] (a33) at 
(\pD,\pB) {};
  \node[dterminal,label=right:{$\kA$, $\kA$-$1$}] (b12) at (\pD,\pE) {};
  \node[dterminal,label=above:{$\kA$, $\kA$-$1$}] (b22) at (\pE,\pD) {};

  \draw[dotted] (\pD,\pA) -- (\pD,\pF);
  \draw[dotted] (\pC,\pA) -- (\pC,\pF);
  \draw[dotted] (\pC,\pD) -- (\pF,\pD);
  \draw[dotted] (\pC,\pC) -- (\pF,\pC);

  \draw[dotted] (\pC,\pB) -- (\pD,\pB);

  \draw[dotted] (\pD,\pE) -- (\pC,\pE);
  \draw[dotted] (\pE,\pD) -- (\pE,\pC);

 \end{tikzpicture}
 \caption{A clause tile}
 \label{fig:clausetile}
\end{center}
\end{figure}
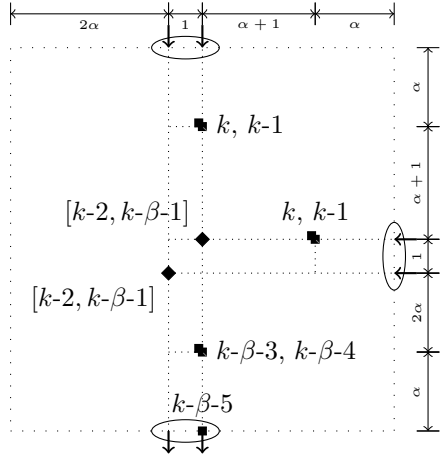

\subsection{Connection Tiles}

In oder to guarantee, that in an optimal solution the proper inputs and outputs 
are connected, we add \emph{connection tiles}.
A horizontal or vertical connection tile enables the connection of a 
horizontal or vertical input with an horizontal or vertical output, respectively 
(see Figure \ref{fig:conn}, left).

\begin{figure}[ht]
\begin{center}
 \begin{tikzpicture}[scale=1]

  \begin{scope}[xshift=0cm, yshift=0cm]
    \tileborder
   \node[terminal] (o1) at (\pA,\pD) {};
   \node[dterminal] (a1) at (\pE,\pD) {};
   \node[terminal] (a2) at (\pD,\pD) {};
   \node[dterminal] (a3) at (\pB,\pD) {};
   \draw (\pA,\pD) -- (\pF,\pD);
   \draw[dotted] (\pA,\pC) -- (\pF,\pC);
   \draw[dotted] (a1) -- ++(0,-\ddelta);
   \draw[dotted] (a2) -- ++(0,-\ddelta);
   \draw[dotted] (a3) -- ++(0,-\ddelta);    

   \node[Steinerpoint] at (\pA,\pD) {};
   \node[Steinerpoint] at (\pB,\pD) {};
   \node[Steinerpoint] at (\pB-0.05,\pD+0.05) {};

   \node[Steinerpoint] at (\pD,\pD) {};
   \node[Steinerpoint] at (\pE,\pD) {};
   \node[Steinerpoint] at (\pE-0.05,\pD+0.05) {};

   \node[Steinerpoint] at (\pF,\pD) {};
  \end{scope}

  \begin{scope}[xshift=4cm, yshift=0cm]
    \tileborder
   \node[terminal] (o1) at (\pA,\pD) {};
   \node[dterminal] (a1) at (\pE,\pD) {};
   \node[terminal] (a2) at (\pD,\pD) {};
   \node[dterminal] (a3) at (\pB,\pD) {};
   \draw[dotted] (\pA,\pD) -- (\pF,\pD);
   \draw (\pA,\pC) -- (\pF,\pC);
   \draw (a1) -- ++(0,-\ddelta);
   \draw (a2) -- ++(0,-\ddelta);
   \draw (a3) -- ++(0,-\ddelta);    
   \draw (o1) -- (\pA,\pC);

   \node[Steinerpoint] at (\pA,\pC) {};
   \node[Steinerpoint] at (\pB,\pC) {};
   \node[Steinerpoint] at (\pD,\pC) {};
   \node[Steinerpoint] at (\pE,\pC) {};
   \node[Steinerpoint] at (\pF,\pC) {};
  \end{scope}

 \end{tikzpicture}
 \caption{Shortest connections for an horizontal connection tile.}
 \label{fig:conn2}
\end{center}
\end{figure}
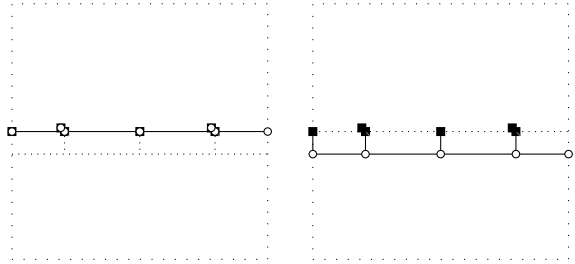

\begin{lemma}
  A minimum connection for a connection tile has length $4\alpha+2$
  if the parity of the input is $1$ and $4\alpha+8$ otherwise.
\end{lemma}
{\it Proof:} 
 The minimum connections are shown in Figure \ref{fig:conn2}.
\endproof 

Note that in a minimum connection the parities of the input and the 
output of a connection tile are the same.

Furthermore, we use \emph{crossing tiles}, each consisting of the union of 
a vertical and an horizontal connection tile (see Figure \ref{fig:conn}, right).
The vertical and horizontal connection of a crossing do not
influence each other, if the depths of the corresponding terminals are
distinct.

\begin{figure}[ht]
\begin{center}
 \begin{tikzpicture}[scale=1.2]
    \tileborder
    \InputI{$k$}
    \OutputI{$k$-$5$}

   \xmarker{\pA}{\pB}{\alpha};
   \xmarker{\pB}{\pD}{\alpha +1};
   \xmarker{\pD}{\pE}{\alpha +1};
   \xmarker{\pE}{\pF}{\alpha};
		
   \ymarker{\pA}{\pB}{\alpha};
   \ymarker{\pB}{\pC}{\alpha};
   \ymarker{\pC}{\pD}{1};
   \ymarker{\pD}{\pE}{\alpha+1};
   \ymarker{\pE}{\pF}{\alpha};
  
   \node[dterminal,label=above:{$k,k$-$1$}] (a1) at (\pE,\pD) {};
   \node[terminal,label=above:{$k$-$2$}] (a2) at (\pD,\pD) {};
   \node[dterminal,label=below:{$k$-$3$,$k$-$4$}] (a3) at (\pB,\pD) {};
   \draw[dotted] (\pA,\pD) -- (\pF,\pD);
   \draw[dotted] (\pA,\pC) -- (\pF,\pC);
   \draw[dotted] (a1) -- ++(0,-\ddelta);
   \draw[dotted] (a2) -- ++(0,-\ddelta);
   \draw[dotted] (a3) -- ++(0,-\ddelta);

  \begin{scope}[xshift=4.3 cm, yshift=0cm]
   \tileborder

   \xmarker{\pA}{\pB}{\alpha};
   \xmarker{\pB}{\pC}{\alpha};
   \xmarker{\pC}{\pD}{1};
   \xmarker{\pD}{\pE}{\alpha+1};
   \xmarker{\pE}{\pF}{\alpha};
   \InputI{}
   \InputII{}
   \OutputI{$k$-$5$}
   \OutputII{$l$-$5$}

   \node[dterminal,label=above:{$k,k$-$1$}] (a1) at (\pE,\pD) {};
   \node[terminal,label=above:{$k$-$2$}] (a2) at (\pD,\pD) {};
   \node[dterminal,label=below:{$k$-$3$,$k$-$4$}] (a3) at (\pB,\pD) {};
   \draw[dotted] (\pA,\pD) -- (\pF,\pD);
   \draw[dotted] (\pA,\pC) -- (\pF,\pC);
   \draw[dotted] (a1) -- ++(0,-\ddelta);
   \draw[dotted] (a2) -- ++(0,-\ddelta);
   \draw[dotted] (a3) -- ++(0,-\ddelta);

   \node[dterminal,label=right:{$l,l$-$1$}] (a1) at (\pD,\pE) {};
   \node[terminal,label=below right:{$l$-$2$}] (a2) at (\pD,\pD) {};
   \node[dterminal,label=right:{$l$-$3$,$l$-$4$}] (a3) at (\pD,\pB) {};
   \draw[dotted] (\pD,\pA) -- (\pD,\pF);
   \draw[dotted] (\pC,\pA) -- (\pC,\pF);
   \draw[dotted] (a1) -- ++(-\ddelta,0);
   \draw[dotted] (a2) -- ++(-\ddelta,0);
   \draw[dotted] (a3) -- ++(-\ddelta,0);

  \end{scope}

 \end{tikzpicture}
\caption{A horizontal connection tile (left) and a crossing 
tile (right).}
\label{fig:conn}
\end{center}
\end{figure}
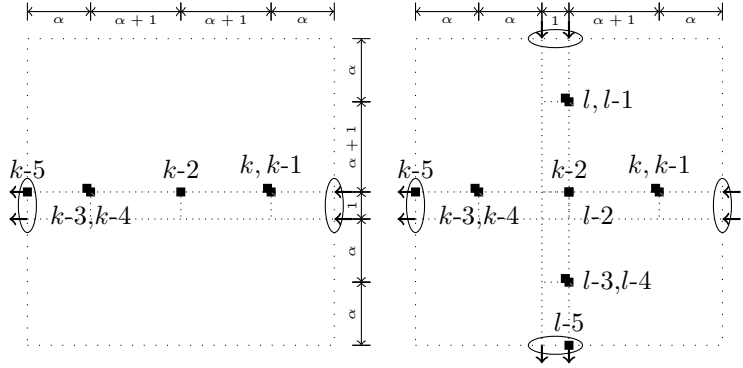


\subsection{Splitter tiles}
Note that a connection tile can only connect inputs and outputs 
that are both at horizontal or both on vertical borders of their tiles and the 
tiles that are connected have to be in the same row
or column of the underlying grid.
But there exist cases where we have to connect one input with two outputs or 
where the input is on a horizontal border and the output is on a vertical borer 
or vise versa. In these cases we ``split'' the path connecting inputs and 
outputs.

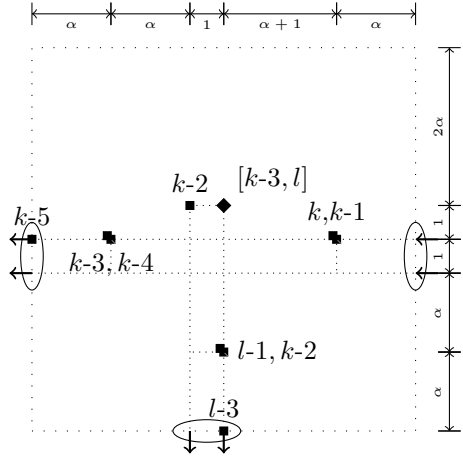
\begin{figure}[ht]
\begin{center}
  \begin{tikzpicture}[scale=1.5]
   \tileborder
   \xmarker{\pA}{\pB}{\alpha};
   \xmarker{\pB}{\pC}{\alpha};
   \xmarker{\pC}{\pD}{1};
   \xmarker{\pD}{\pE}{\alpha+1};
   \xmarker{\pE}{\pF}{\alpha};
  
   \ymarker{\pA}{\pB}{\alpha};
   \ymarker{\pB}{\pC}{\alpha};
   \ymarker{\pC}{\pD}{1};
   \ymarker{\pD}{\pDD}{1};
   \ymarker{\pDD}{\pF}{2\alpha};

\InputI{$k$};
\OutputI{$\kA$-$5$};
\OutputII{$\kB$-$3$};

   \node[cascade,label=right:{$[\kA$-$3,\kB]$}] (cc1) at
       (\pD,\pDD) {};
   \node[terminal,label=above:{$\kA$-$2$}] (e) at (\pC,\pDD) {};
   \node[dterminal,label=above:{$\kA$,$\kA$-$1$}] (a2) at (\pE,\pD) 
     {};
   \node[dterminal,label=below:{$\kA$-$3,\kA$-$4$}] (b1) at (\pB,\pD) {};
   \node[dterminal,label=right:{$\kB$-$1,\kA$-$2$}] (c1) at (\pD,\pB) {};
   

   \draw[dotted] (\pA,\pC) -- (\pF,\pC);
   \draw[dotted] (\pA,\pD) -- (\pF,\pD);
   \draw[dotted] (\pC,\pDD) -- (\pC,\pA);
   \draw[dotted] (\pD,\pDD) -- (\pD,\pA);
   \draw[dotted] (cc1) -- ++(-\ddelta,0);
   \draw[dotted] (c1) -- ++(-\ddelta,0);
   \draw[dotted] (a2) -- ++(0,-\ddelta);
   \draw[dotted] (b1) -- ++(0,-\ddelta);

 \end{tikzpicture}
 \caption{Horizontal splitter tile: position of the terminals and the terminal 
cascade 
and their depths.}
  \label{fig:splittertile}
\end{center}
\end{figure}

To this end, we introduce \emph{splitter tiles} (see Figure 
\ref{fig:splittertile}).
A splitter tile contains one input with depth $k$ and two outputs.
It is designed in such a way that in an optimal solution the parities of all 
inputs and outputs are the same. 
There are two types of splitter tiles; one, where the input is at the upper 
border and one where the input is on the right border. 
As these types of tiles are the same up to symmetry, we restrict ourselves to 
define 
and analyze splitter tiles where the input is on the right border of the tile.
The terminal cascade plays a crucial role here.

\begin{proposition}
 Each feasible connection of a splitter tile with a terminal cascade containing 
$\gamma$ terminals has length at least $6\alpha+\gamma$.
\end{proposition}
{\it Proof:} 
 Consider an horizontal splitter tile $t$ and denote by $C$ the terminals of 
the 
 terminal cascade. As $t$ contains one input and two  outputs, the induced 
 length is at least $6\alpha$ by Prop. \ref{prop:doubleterminals}.
 Let $S$ be the set of Steiner points that are placed on the position of the 
 terminal cascade. If $S$ is empty, then the distance between each terminal of 
 $C$ and its parent is at least $1$, thus the total length of  the 
 connection is at least $6\alpha+|C|=6\alpha+\gamma$.

 If on the other hand $S\neq\emptyset$, then let $s\in S$ be a vertex with 
 highest depths. Consider the subtree rooted at $s$. As all terminals of $C$ 
 have distinct depths and by the observation that the double terminals of 
 the input cannot be in the subtree, there must be a vertex in the subtree 
 outside the tile. 
 But then the induced length of $t$ is at least $8\alpha>4\alpha + \gamma$, 
 where   $6\alpha$ comes from the double terminals and $2\alpha$ from the 
 subtree rooted at $s$.
\endproof 

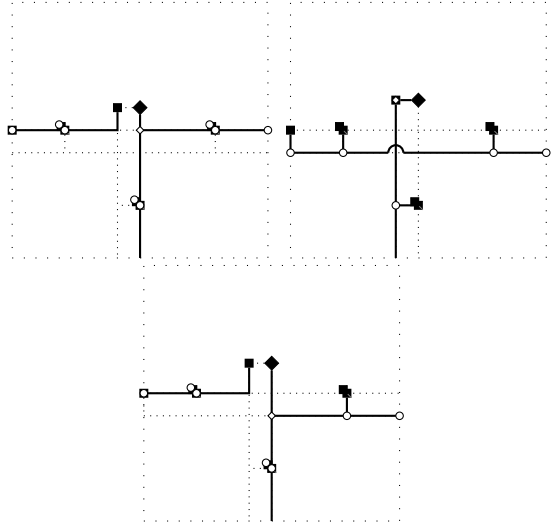
\begin{figure}[ht]
\begin{center}
\begin{tikzpicture}

   \begin{scope}[xshift=3.1cm,yshift=0.9cm, scale=1]
    \tileborder
    \node[cascade] (cc1) at (\pD,\pDD) {};
    \node[terminal] (e) at (\pC,\pDD) {};
    \node[dterminal] (a2) at (\pE,\pD) {};
    \node[dterminal] (b1) at (\pB,\pD) {};
    \node[dterminal] (c1) at (\pD,\pB) {};
    \node[terminal] (d) at (\pA,\pD) {};

    \draw[dotted] (\pA,\pC) -- (\pF,\pC);
    \draw[dotted] (\pA,\pD) -- (\pF,\pD);
    \draw[dotted] (\pC,\pDD) -- (\pC,\pA);
    \draw[dotted] (\pD,\pDD) -- (\pD,\pA);
    \draw[dotted] (cc1) -- ++(-\ddelta,0);
    \draw[dotted] (c1) -- ++(-\ddelta,0);
    \draw[dotted] (a2) -- ++(0,-\ddelta);
    \draw[dotted] (b1) -- ++(0,-\ddelta);

   \draw[thick] (\pF,\pD) -- (\pD,\pD);
   \draw[thick] (cc1) -- (\pD,\pA);
   \draw[thick] (e) -- ++(0,-\ddelta) -- (\pA,\pD);

   \node[Steinerpoint] at (b1) {};
   \node[Steinerpoint] at (a2) {};
   \node[Steinerpoint] at (c1) {};
   \node[Steinerpointcascade] at (\pD,\pD) {};
   \node[Steinerpoint] at (\pF,\pD) {};
   \node[Steinerpoint] at (\pA,\pD) {};

    \node[Steinerpoint] at (\pE-0.075,\pD+0.075) {};
    \node[Steinerpoint] at (\pB-0.075,\pD+0.075) {};
    \node[Steinerpoint] at (\pD-0.075,\pB+0.075) {};

  \begin{scope}[xshift=3.7cm, yshift=0cm]
    \tileborder
    \node[cascade] (cc1) at (\pD,2*\ddelta) {};
    \node[terminal] (e) at (\pC,2*\ddelta) {};
    \node[dterminal] (a2) at (\pE,\pD) {};
    \node[dterminal] (b1) at (\pB,\pD) {};
    \node[dterminal] (c1) at (\pD,\pB) {};
    \node[terminal] (d) at (\pA,\pD) {};

    \draw[dotted] (\pA,\pC) -- (\pF,\pC);
    \draw[dotted] (\pA,\pD) -- (\pF,\pD);
    \draw[dotted] (\pC,2*\ddelta) -- (\pC,\pA);
    \draw[dotted] (\pD,2*\ddelta) -- (\pD,\pA);
    \draw[dotted] (cc1) -- ++(-\ddelta,0);
    \draw[dotted] (c1) -- ++(-\ddelta,0);
    \draw[dotted] (a2) -- ++(0,-\ddelta);
    \draw[dotted] (b1) -- ++(0,-\ddelta);

    \draw[thick] (\pF,\pC) -- (\pC+0.1,\pC);
    \draw[thick] (\pC+0.1,\pC) arc (0:180:0.1cm);
    \draw[thick] (\pC-0.1,\pC) -- (\pA,\pC);
    \draw[thick] (a2) -- ++(0,-\ddelta);
    \draw[thick] (b1) -- ++(0,-\ddelta);
    \draw[thick] (cc1) -- (e) -- (\pC,\pA);
    \draw[thick] (c1) -- ++(-\ddelta,0);
    \draw[thick] (d) -- ++(0,-\ddelta);

    \node[Steinerpoint] at (\pB,\pC) {};
    \node[Steinerpoint] at (\pE,\pC) {};
    \node[Steinerpoint] at (\pC,\pB) {};
    \node[Steinerpointcascade] at (\pC,2*\ddelta) {};
   \node[Steinerpoint] at (\pF,\pC) {};
   \node[Steinerpoint] at (\pA,\pC) {};

  \end{scope}

  \begin{scope}[xshift=1.75cm, yshift=-3.5cm]
    \tileborder
    \node[cascade] (cc1) at (\pD,2*\ddelta) {};
    \node[terminal] (e) at (\pC,2*\ddelta) {};
    \node[dterminal] (a2) at (\pE,\pD) {};
    \node[dterminal] (b1) at (\pB,\pD) {};
    \node[dterminal] (c1) at (\pD,\pB) {};
    \node[terminal] (d) at (\pA,\pD) {};

    \draw[dotted] (\pA,\pC) -- (\pF,\pC);
    \draw[dotted] (\pA,\pD) -- (\pF,\pD);
    \draw[dotted] (\pC,2*\ddelta) -- (\pC,\pA);
    \draw[dotted] (\pD,2*\ddelta) -- (\pD,\pA);
    \draw[dotted] (cc1) -- ++(-\ddelta,0);
    \draw[dotted] (c1) -- ++(-\ddelta,0);
    \draw[dotted] (d) -- ++(0,-\ddelta);

    \draw[thick] (\pD,\pC) --  (\pF,\pC);
    \draw[thick] (cc1) -- (\pD,\pA);
    \draw[thick] (e) -- (\pC,\pD) -- (\pA,\pD);
    \draw[thick] (a2) -- ++(0,-\ddelta);

    \node[Steinerpoint] at (\pB,\pD) {};
    \node[Steinerpoint] at (\pE,\pC) {};
    \node[Steinerpoint] at (\pD,\pB) {};
    \node[Steinerpointcascade] at (\pD,\pC) {};

   \node[Steinerpoint] at (\pF,\pC) {};
   \node[Steinerpoint] at (\pA,\pD) {};

    \node[Steinerpoint] at (\pB-0.075,\pD+0.075) {};
    \node[Steinerpoint] at (\pD-0.075,\pB+0.075) {};

  \end{scope}
  \end{scope}
   
 \end{tikzpicture}
 \caption{The three possible shortest connections for a splitter tile.}
 \label{fig:splittertile2}
\end{center}
\end{figure}

Now we analyze the length of shortest connections for a splitter tile:

\begin{lemma}\label{lemma:splitter}
 Let $A$ be a minimum connection for a splitter with a terminal cascade 
 containing $\gamma$ terminals and set $L=6\alpha+\gamma+3$.
 If the parities of the inputs and outputs are $1$, then $A$ has length $L$.
 If the parities of the inputs and outputs are $0$, then $A$ has length $L+8$.
 Finally, if the parity of the input is $0$ and the parities of the outputs are 
 $1$ then $A$ has length $L+1+2\gamma$.
\end{lemma}
{\it Proof:} 
  Figure \ref{fig:splittertile2} shows the three possible shortest
  connections. 
  In the first two cases each terminal of the terminal cascade 
  has to be connected to the tree by an edge of length $1$, but in the 
  last case, they are connected by edges of length $2$.
\endproof 

 In the third case of Lemma \ref{lemma:splitter} the parity swit\-ches from $0$ 
 at the input to $1$ at the outputs, that is, the corresponding literal switches
 from {\false} to  {\true}.
 We call this type of connection a \emph{forbidden connection}.
 As this is not allowed we have to ensure that such connections are never a
 part of an optimal solution. 
 So we assume that the following assumption is satisfied during the remainder 
 of the paper. 
\begin{assumption}\label{ass2}
 If $I$ is an instance with $k$ double terminals and $l$ splitter tiles with 
 terminal cascades containing $\gamma_1,\ldots, \gamma_l$ terminals, 
 respectively, then  an optimal solution hast cost less than $2k\alpha + 
\sum_{i\in\{1,\ldots,l\}} \gamma_i + \min_{i\in\{1,\ldots, l\}} \gamma_i$.
\end{assumption}

If this assumption is satisfied, then the solution does not contain forbidden 
connections.

\subsection{Additional connections}
 Using the variable, clause, splitter and connection tiles we are now able to 
 build the main part of our instance. But there are still some ``open'' outputs 
that are not
 connected yet, for example the outputs of the clauses.
 Thus we have to add terminals with  consecutive depths in order to connect
 these outputs to the root $r$. For every additional input and output that we 
 use we add a double terminal in order to coincide with the structure of the 
 prototype tile.
 If two such connections meet, we add a terminal cascade such that these
 connections can merge into a single vertex. Finally, we add a terminal cascade 
at 
 the root in order to guarantee the existence of a feasible solution.
 Note that the minimum total length of the edges required to connect these 
additional terminals
 is always the same.

\subsection{Realizations of truth assignments and NP-com\-ple\-te\-ness}
\label{section:values}

Let $T$ be the set of all tiles. For each tile $t\in T$ we denote by $L(t)$ the 
minimum
length of a shortest connection for $t$. Then $L:=\sum_{t\in T} L(t)$ is a
lower bound for the length of a feasible solution.

Let $\pi:\{x_1,\ldots,x_n\}\rightarrow \{0,1\}$ be a truth assignment.
We construct a feasible solution by setting the parities of the variable
tiles according to the truth assignment, inducing the parities of all inputs
and outputs of the tiles and
call the resulting arborescence a \emph{realization} of $\pi$.

\begin{figure}[ht]
  \begin{center}

   \begin{tikzpicture}[scale=0.29]
     \input{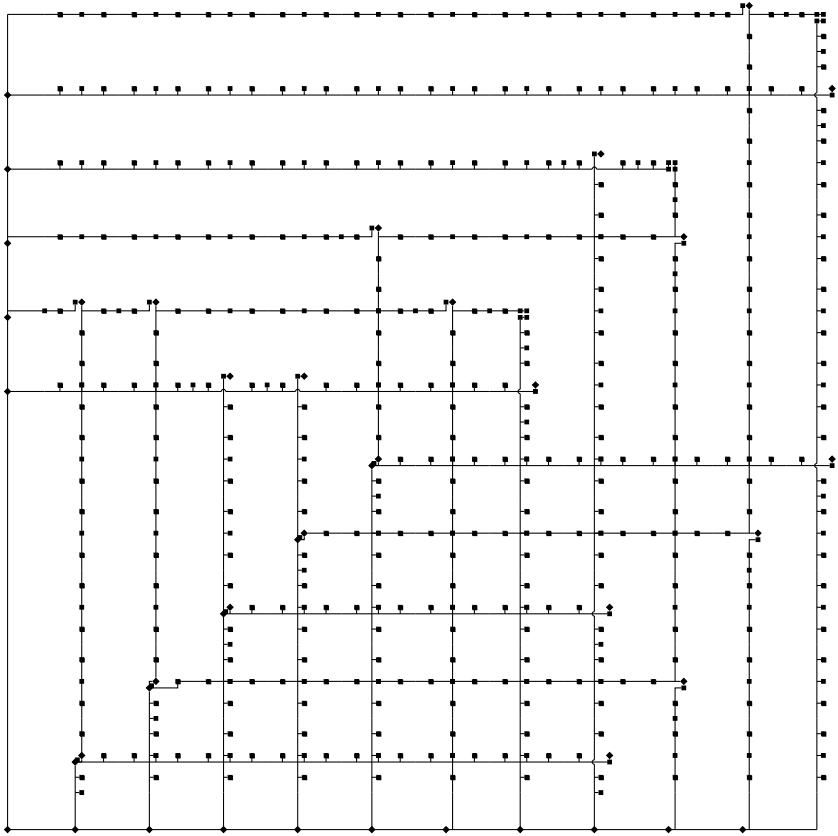}
   \end{tikzpicture}

   \caption{An instance of the rectilinear Steiner arborescence with depth 
            restriction problem and a shortest solution for it.}
   \label{fig:instance}
  \end{center}
\end{figure}

Figure \ref{fig:instance} shows a transformed instance for the 
\emph{Max-2-Sa} instance $(\mathcal{V},\mathcal{C})$
defined by  $V=\{x_1,x_2,x_3\}$, $\mathcal{C}=\{C_1,\ldots,C_5\}$,
  $C_1=\{x_1,x_2\}, 
  C_2=\{x_1,\overline{x_2}\}, 
  C_3=\{\overline{x_1}, x_2\},$ 
  $C_4=\{\overline{x_1}, x_3\}$ and 
  $C_5=\{\overline{x_2}, \overline{x_3}\}$.
Moreover, a shortest solution corresponding to the truth assignment 
$x_1=x_3=${\,\true} and
$x_2=${\,\false} is shown. 
The given depths have been omitted for clarity of presentation.
The figure also illustrates, how the open outputs are connected to the root.

Next, we analyze the length of a realization.
As seen in Lemma~\ref{lemma:clause} the induced connection of a clause tile has 
length 
$6\alpha+9$ or $6\alpha+10$ if the clause is satisfied under $\pi$ and 
$6\alpha+11+\beta$ 
otherwise.
For all other tiles, the length of the induced connection is at most $8$ units
longer than a minimum connection for that tile.
Let $u$ be the number of clauses that are not satisfied under $\pi$. 
Then the length $L(\pi)$ of the solution is at least $L+2u\beta$ and at 
most $L+2u\beta + 3N^2 < L+3u\beta$.

For the length $\ell$ of a realization satisfying $n-u$ of the clauses, we have
\begin{equation}\label{eq:beta}
 \ell \in [L + u\beta, L + u\beta + 10(nm)^2] =: B_u.
\end{equation}

Now the values for $\alpha, \beta $ and $\gamma$ can be specified.
We have to set $\beta$ such that the length of a shortest realization 
indicates, how many clauses are satisfied. To this end, 
the sets $B_u$, $u\in\{1,\ldots, m\}$, have to be distinct and we set
\begin{equation}
 \beta = 20(nm)^2.
\end{equation}

In order to satisfy Assumption \ref{ass2}, we observe that every feasible
realization has length at most
\begin{equation}\label{eq:l1}
 2l\alpha + \sum_{i\in\{1,\ldots, l\}} \gamma_i + 10(nm)^2 + m\beta,
\end{equation}
where $\gamma_i$ is the number of terminals in the terminal cascade of the 
$i$'s splitter tile.
Thus Assumption \ref{ass2} is satisfied if
\begin{equation}\label{eq:l2}
 4(nm)^3 > \gamma_i > (nm)^3
\end{equation}
for all $i\in\{1,\ldots, l\}$.
This can be achieved by setting the depths of the four terminals of each
variable tile to $4(nm)^3$ and the depths of the connection and splitter tiles 
appropriately. This is possible as every root-terminal-path passes at most two 
splitter tiles.

Using (\ref{eq:l1}) in (\ref{eq:l2}), we conclude, that the length of a
realization for an instance with $k$ double terminals is at most
\begin{equation}
 2k\alpha + 4l(nm)^3 +10(nm)^2+ m20(nm)^2
\end{equation}
which is at most $ 2k\alpha + (nm)^4$ if $nm$ is sufficiently large.
By setting
 $\alpha:= (nm)^4$
 Assumption \ref{ass1} is satisfied.
Note that the values for  $\alpha, \beta$ and $\gamma$ and the number of 
terminals are polynomially bounded in $n+m$. Thus we have a polynomial 
transformation.

All the previous observations together give us the main result of this paper.

\begin{theorem}
 The depth-restricted rectilinear Steiner arborescence problem is strongly 
NP-com\-ple\-te.
\end{theorem}
{\it Proof:} 
 The problem is obviously in NP.
 Using the transformation described in this paper we can transform a 
\emph{Max-2-Sat} instance $(\mathcal{V},\mathcal{C})$ into an instance 
 $I$ of the depth-restricted rectilinear Steiner arborescence problem.
 As the number of terminals, the depths and the distances in $T$ are
 polynomially bounded in $|\mathcal{V}|+|\mathcal{C}|$, the transformation is
 a polynomial one.
 We conclude that as \emph{Max-2-Sat} is strongly NP-complete, so is the 
depth-restricted rectilinear Steiner arborescence
 problem.
\endproof 

\bibliography{massberg_bib}{}
\bibliographystyle{plain}

\end{document}